# Adaptive Real-Time Software Defined MIMO Visible Light Communications using Spatial Multiplexing and Spatial Diversity

Peng Deng and Mohsen Kavehrad
Department of Electrical Engineering
The Pennsylvania State University, University Park, PA, United States
E-mail: {pxd18, mkavehrad}@psu.edu

*Abstract*—In this paper, we experimentally demonstrate a real-time software defined multiple input multiple output (MIMO) visible light communication (VLC) system employing link adaptation of spatial multiplexing and spatial diversity. Real-time MIMO signal processing is implemented by using the Field Programmable Gate Array (FPGA) based Universal Software Radio Peripheral (USRP) devices. Software defined implantation of MIMO VLC can assist in enabling an adaptive and reconfigurable communication system without hardware changes. We measured the error vector magnitude (EVM), bit error rate (BER) and spectral efficiency performance for single-carrier M-QAM MIMO VLC using spatial diversity and spatial multiplexing. Results show that spatial diversity MIMO VLC improves error performance at the cost of spectral efficiency that spatial multiplexing should enhance. We propose the adaptive MIMO solution that both modulation schema and MIMO schema are dynamically adapted to the changing channel conditions for enhancing the error performance and spectral efficiency. The average error-free spectral efficiency of adaptive 2x2 MIMO VLC achieved 12 b/s/Hz over 2 meters indoor dynamic transmission.

*Keywords—Visible Light Communication; MIMO; Spatial multiplexing; Spatial diversity; Link Adaptation; Software Defined Radio; Light-Emitting Diodes*

## I. INTRODUCTION

The worldwide growth in wireless mobile data traffic has led to the development of new technologies for high capacity and energy efficient wireless communication systems. This fact results in an increasing throughput requirement from the next generation mobile communication networks (5G), which are expected to address several critical challenges, such as broadband capacity, spectral efficiency, power efficiency, quality-of-services and mobility coverage. Visible light communication system (VLC) has been accepted as part of the 802.15.7 task group and proposed as a supplement technology in 5G networks standards. VLC based on light-emitting diodes (LEDs) merges lighting and data communications in applications due to their energy efficiency, spectral efficiency, security and reliability[1]. The lighting white LEDs have the advantage of long life time, energy-efficiency and environmental friendliness. VLC using lighting LEDs could provide many advantages, such as no electromagnetic interference (EMI), integration with indoor lighting, worldwide available and unlicensed bandwidth[2]. The VLC application as shown in Fig.1 is relevant to aerospace, automotive, healthcare industries, 5G networks, WPANs, Internet-of-Things, and ad-hoc networks, as it focuses on development of RF-interference-free high-bandwidth communication networks using lighting white LEDs.

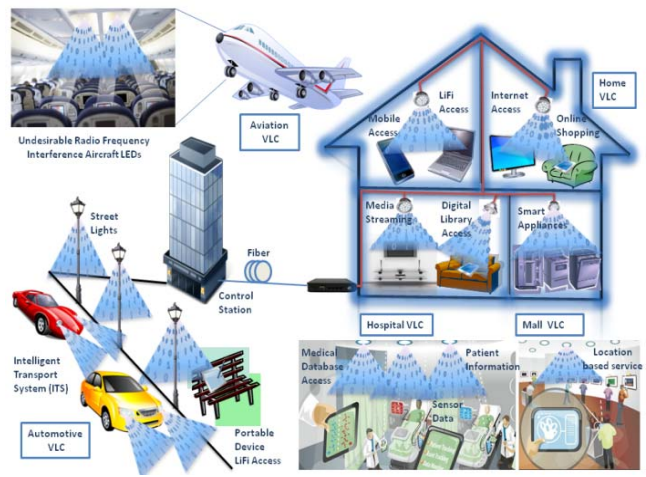

Fig. 1. Visible Light Communications Applications

The commercial white-light LED utilized for general lighting is mostly based on a blue LED chip covered by a phosphor layer, due to the relatively lower cost and complexity compared to red-green-red (RGB) LEDs. However, the phosphorescent component would limit the modulation bandwidth of phosphor-LED. To enhance modulation bandwidth of phosphor-LED, several technologies have been proposed, such as utilizing a blue filter and using analogue/digital equalization [3]. While the blue filtering introduces high signal attenuation, which would reduce the longest transmission distance in VLC.

High spectrally efficient modulation format is a straightforward way to increase the capacity under the limited bandwidth, such as orthogonal frequency-division multiplexing (OFDM), carrier-less amplitude and phase (CAP) modulation and Nyquist single carrier modulation [4, 5]. OFDM can maximize its transmission capacity by bit- and power-loading under the constraint of a required bit error rate (BER). However, OFDM has higher peak-to-average-power ratio (PAPR), which is crucial for the nonlinear modulation characteristics of LEDs in VLC systems [6, 7]. Most CAP systems are designed and implemented in the digital domain by applying a relatively high-rate and high-resolution A/D . Also, CAP systems suffer from higher sampling jitter sensitivity. Despite the complexity of adding analog multipliers, QAM architecture has the benefits of less sensitivity to the nonlinearity and sampling jitter as well as lower sampling rate and resolution A/Ds required for real-time practices.

Moreover, indoor Illumination system is typically provided by large numbers of lighting lamps in a room and each lamp is made up of multiple LEDs. Thus, LEDs based VLC naturally constitutes a MIMO system, which can achieve parallel transmission of independent data[8] and tolerance to shadowing. There are several MIMO transmission techniques, such as Spatial Diversity, Spatial Multiplexing and Spatial Modulation[9]. Reference [10] demonstrated a 4x9 spatial multiplex MIMO Visible-Light Communications using 16-QAM OFDM with each transmitting signals at 250 Mb/s in 1 m range. 4×4 spatial multiplex MIMO VLC is demonstrated by OOK format at 50 Mb/s 2 m range[11]. Demonstration of 2 x 2 MIMO VLC is achieved at 500 Mb/s in 40 cm range by using 4-QAM Nyquist Single Carrier With Frequency Domain Equalization[12]. Spatial-Diversity MIMO VLCs are proposed by using fisheye lens wide filed-of-view receiver [13, 14].

However, currently most MIMO VLC demonstrations feature offline signal processing, fixed MIMO schema and static ideal channel condition estimation. In practical indoor wireless mobile communications, the complex channel conditions are expected to be changed dramatically and the spectral efficiency must be optimized adaptively. For example, MIMO VLC system using fixed spatial multiplexing will suffer link interruption, once any sub-channel condition becomes degraded. Real time data processing on field programmable gate arrays (FPGAs) based software-defined device can build a fully real time reconfigurable and adaptive optical wireless system by upgrading the software. The network architecture should also be designed to support adaptive MIMO schemes with distributed smart antenna systems. Thus, a real-time adaptive MIMO optical wireless system is necessary to deal with the dynamic complex channel conditions and enhance the robust mobility and link reliability in future wireless mobile networks.

In this paper, we experimentally demonstrate an adaptive real-time software-defined Single-Carrier M-QAM MIMO visible light communication system by using spatial multiplexing and spatial diversity. The system uses two independent phosphorescent white LED transmitters with 10MHz bandwidth in the absence of blue filters. Single-carrier M-QAM is less sensitive to LED's nonlinearity compared with high PAPR OFDM. QAM also has lower sampling rate and resolution A/Ds required than CAP. Real-time MIMO signal processing is implemented by using FPGA based Universal Software Radio Peripheral devices. Software defined implantation of MIMO VLC can assist in enabling an adaptive and reconfigurable communication system without hardware changes. We measured the error vector magnitude (EVM), bit error rate (BER) and spectral efficiency performance for single-carried M-QAM MIMO VLC using spatial diversity and spatial multiplexing. Results show that spatial diversity MIMO VLC improves the error performance at the cost of spectral efficiency that spatial multiplexing should enhance. We propose the adaptive MIMO solution that both modulation schemas and MIMO schemas are dynamically adapted to the changing channel conditions for enhancing the error performance and spectral efficiency. The average error-free spectral efficiency of adaptive 2x2 MIMO VLC achieved 12 b/s/Hz over 2 meters indoor dynamic transmission.

## II. SYSTEM MODEL

Let us consider an indoor visible light communication system with $M$ transmitting antennas (LEDs) and $N$ receiving antennas (photo-detectors). The input data $X$ is multiplexed into $M$ parallel transmit data streams $x_j$ (j = 1,..., $M$). Each of the new data streams are used to intensity modulate a light source. When transmitting two signals by modulating them with QAM, the transmitted signal will be of the form:

$$x(t) = I(t)\cos(2\pi f_c t) - Q(t)\sin(2\pi f_c t) \quad (1)$$

where $I(t)$ and $Q(t)$ are the modulating signals, $f_c$ is the carrier frequency. As all data streams are transmitted at the same time, the received signal is a linear combination of all $x_j$. Its discrete-time baseband MIMO channel model can be represented as

$$\mathbf{y} = \mathbf{H}\mathbf{x} + \mathbf{n}, \quad (2)$$

where $\mathbf{n}$ is the sum of ambient shot light noise and thermal noise with zero mean and a variance $\sigma^2$, and the noise power spectral density is $N_0$. It is assumed that $N_t$ data symbols $x_1$, $x_2$, … , $x_{Nt}$ are chosen randomly, equally-likely and independently to form an input data vector $\mathbf{x} = [x_1, x_2, ..., x_{N_t}]^T \in \mathbf{A}^M$ where $A$ is a given modulation constellation. Each Tx transmits independent symbols of the same power: $<\mathbf{x}\mathbf{x}^*> = P_t/N_t \cdot \mathbf{I}$, where $P_t$ is the total Tx power, $\mathbf{I}$ is the identity matrix, $<>$ and * denote expectation and Hermitian conjugation respectively. The received signal vector is denoted by $\mathbf{y} = [y_1, y_2, ..., y_{N_r}]^T$. Notation $\mathbf{H}$ denotes an $N_t \times N_r$ channel matrix, in which $h_{nm}$ is the non-negative and real coefficient between the $m^{th}$ transmitter antenna and the $n^{th}$ receiver antenna.

In order to de-multiplex the signals and retrieve the transmitted data, the MIMO system has first to estimate the channel estimation coefficients between a pair of Tx and Rx. To achieve this, training sequences of binary phase shift keying signals are periodically inserted in front of the data streams to obtain the matrix for channel estimation. Channel estimation should consider four different channels. Each row of the 2D array contains two training sequences, one from each transmitter.

$$\mathbf{t} = \begin{pmatrix} ts_1 & 0 \\ 0 & ts_2 \end{pmatrix} \quad (3)$$

Thus, perform the channel estimation on each training sequence in each row then determine the four channel estimates. On receiving all pilot signals $\mathbf{y}_t$, an $\mathbf{H}$ matrix detailing channel estimation is logged. It is only then that the

Tx can proceed with the simultaneous transmission of data on all channels.

$$\mathbf{H} = \frac{\mathbf{y_t}}{\mathbf{t}} = \begin{pmatrix} y_{11}/ts_1 & y_{12}/ts_2 \\ y_{21}/ts_1 & y_{22}/ts_2 \end{pmatrix} \quad (4)$$

The simplest method to estimate the transmitted data would be to invert **H** and multiply it with the received vector known as the Zero-forcing ZF [15]:

$$\mathbf{W} * \mathbf{y} = \mathbf{x}_{est} + \mathbf{n}, \quad (5)$$

where $W$ is the beam former $\mathbf{H}^{-1}$. However, if the values of **H** are low, the noise vector **n** increases leading to noise amplification. If **H** is rank deficient, then matrix inversion cannot be performed. In such cases, the pseudo inverse of **H** can be used given by:

$$\mathbf{H}^{\dagger} = (\mathbf{H}^*\mathbf{H})^{-1}\mathbf{H}^*, \quad (6)$$

where $\mathbf{H}^*$ is the conjugated transpose of **H**. An estimate of **x** can then be made by using this pseudo inverse Zero-forcing, the de-multiplexing and post-equalization can be simultaneously realized. Assuming that the channel matrix is perfectly known to the receiver via a training sequence, and that the signals from individual LEDs are independent and equi-powered [15], then the spectral efficiency of the MIMO channels is given by

$$S = \log_2\left[\det\left(\mathbf{I}_N + \frac{P_t}{N_t\sigma_0^2}\mathbf{H}\mathbf{H}^*\right)\right]$$
$$= \sum_{i=1}^{N_m} \log_2\left(1 + \frac{P_t}{N_t\sigma_0^2}\lambda_i^2\right) \quad (7)$$
$$\leq \log_2\left(1 + \frac{P_t}{N_t\sigma_0^2}\|\mathbf{H}\|^2\right)$$

where $N_m = \min(N_t, N_r)$, $\boldsymbol{\lambda} \triangleq [\lambda_1^2,...,\lambda_{N_m}^2]^T$ is eigen value vector of $\mathbf{H}\mathbf{H}^*$, $\|\;\|$ denotes the Frobenius norm, $\|\mathbf{H}\|^2 = \sum_{i=1}^{N}\sum_{j=1}^{M}|h_{i,j}|^2$.

The adaptive MIMO is mainly based on the approximated expression for the average bit error-rate (BER) of $M$-ary QAM symbols and channel capacity of MIMO system. An approximate expression of average bit error probability of $M$-ary square QAM can be obtained by

$$P_b \cong \frac{2(\sqrt{M}-1)}{\sqrt{M}\log_2^M}\mathrm{erfc}\left(\sqrt{\frac{3}{2(M-1)}\gamma}\right) \quad (8)$$

where $\gamma$ is the received signal to noise ratio (SNR) per symbol, and $\mathrm{erfc}(x)$ is the complementary error function. The analytical error-probability bound for an AWGN SISO channel with $M$-QAM modulation is given by [16]

$$P_b \leq \frac{1}{5}\exp(-1.5\frac{\gamma}{M-1}) \quad (9)$$

The maximum spectral efficiency for a given $BER_{tgt}$ can be obtained by

$$s = \log_2(1+K\gamma) \quad (10)$$

where $K = \dfrac{-3}{2\ln(5BER_{tgt})}$.

The adaptive MIMO system can be defined by the constrained optimization problem.

$$\max_{MIMO, s_i, P_i} S = E_\lambda[\sum_{i=1}^{N_m} s_i(\boldsymbol{\lambda})] \quad (11)$$

subject to

$$E_\lambda[\sum_{i=1}^{N_m} P_i(\boldsymbol{\lambda})] \leq P_t \quad (12)$$
$$BER_i(\boldsymbol{\lambda}) \leq BER_{tgt} \quad (13)$$
$$M_i(\boldsymbol{\lambda}) \in \{M_0, M_1,...,M_n\} \quad (14)$$
$$MIMO(\boldsymbol{\lambda}) \in \{SM, SD\} \quad (15)$$
$$P_i(\boldsymbol{\lambda}) \geq 0, s_i(\boldsymbol{\lambda}) \geq 0, i=1,...,N_m \quad (16)$$

Equation (11) expresses the optimization strategy that maximizing the spectral efficiency by adapting the *MIMO* scheme, transmission rate $s_i$ and power $P_i$ in each subchannel. Equation (12) shows that the average total transmit power cannot exceed $P_t$ and (13) requires that at any instant the BER in any subchannel must remain below a predetermined target level $BER_{tgt}$. We consider more practical discrete rate systems in (14), where we restrict the available constellation size $M_i$ to a finite set of integers instead of any nonnegative real values. In particular, we use only square QAMs as the available component modulation schemes; i.e., $M_i = 4^i$, $i = 0, 1, ..., N$. Based on the conditions of MIMO channel estimations, we can adapt the MIMO schemes in (15) such as spatial multiplexing (SM) and spatial diversity (SD) to maximize the achievable spectral efficiency. Equation (16) is the nonnegative constraint on power and rate. The expectation over $\boldsymbol{\lambda}$ in (12) implies that we assume the random process $\boldsymbol{\lambda}$ is ergodic. These enable the MIMO adaptation in the time domain and spatial domain.

III. EXPERIMENTAL SETUP

Figure 2 presents a block diagram and experimental setup of this 2X2 M-QAM adaptive MIMO VLC system. The random binary data is generated in Labview and would be first split into two parallel streams, one for each transmitter (TX) channel. We develop adaptive MIMO modes control modules, which can adjust the optimal modulation formats and MIMO schemes to maximize the spectral efficiency and error performance according to the real-time MIMO channel conditions. There are two types of MIMO schemes for selection: Spatial Diversity(SD) MIMO and Spatial Multiplexing(SM) MIMO. The former one requires each Tx transmit the same data stream to improve the antenna array gain, while the latter one needs each LED send the different parallel streams to enhance the spectral efficiency gain. In each channel, the bit stream is mapped into $M$-ary quadrature amplitude modulation ($M$-QAM) and there are four types of modulation formats: 4-QAM,16-QAM,64-QAM and 256-QAM.Thus, there are eight adaptive MIMO modes available for link adaptation(SM-4,SM-16,SM-64,SM-256,SD-4,SD-16,SD-64,SD-256).Initilaly, we choose the default mode of SM-64(spatial multiplexing 64-QAM). Then channel training sequences are inserted into the symbol streams. After adding cyclic prefix (CP) and up-sampling, pulse shaping by a raised-cosine filter is employed.

The host computer transmits data streams to Universal Software Radio Peripheral (USRP) X310 via the 10G SFP Ethernet interfaces. After digital to analog conversion (DAC), the complex QAM streams are up-converted to RF frequency and generate the real analogue RF signals. The up-conversion used here provide flexible frequency allocation and offer RF frequency for I/Q modulation. At the analogue transmitter front-end, the output signal was first amplified and then superimposed onto the LED bias current by the aid of a bias-tee. The light source was a phosphorescent white LED that consists of four chips, each providing a luminous flux of 520 lm with a 120° full opening angle when driving at 700 mA. An aspheric lens was fixed to make sure the light transmits along the 2 meter straight direction. Light from the white LED was imaged onto a high-speed PIN photo-detector through an aspheric convex lens. The photocurrent signal was amplified by a low noise trans-impedance amplifier (TIA). A first-order analogue post-equalization was developed to extend the bandwidth based on phosphorescent white LED.

Subsequently, the received signals from each of the two receivers are routed to a RX channel of USRP X310. The X310 are high-performance, scalable software defined radio (SDR) platforms for designing and deploying next generation wireless communications systems. The hardware architecture combines two extended bandwidth daughterboard slots, dual high-speed SFP(+) interface ports for 1/10 Gigabit Ethernet, 200 MS/s ADC, 800 MS/s DAC, and a large user-programmable Kintex-7 FPGA in a convenient desktop to provide the best-in-class real-time hardware performance.

After down-converting to baseband and down sampling by a matched filter, the training sequences are acquired for synchronization and channel estimation. After removing CP, the received data streams are processed in a MIMO de-multiplexer. The final streams are then passed through the QAM demodulator to recover the original binary stream. Based on the channel estimation conditions, the adaptive MIMO modes selection module will calculate and select the corresponding optimum modes including modulation formats and MIMO schemes. We then update the three bit binary code for adaptive MIMO modes and feedback the adaptive mode code to the transmitter by using RF uplink. In Labview, we measured and analyzed the constellation diagram, EVM, BER and spectral efficiency performance for adaptive $M$-QAM MIMO VLC system.

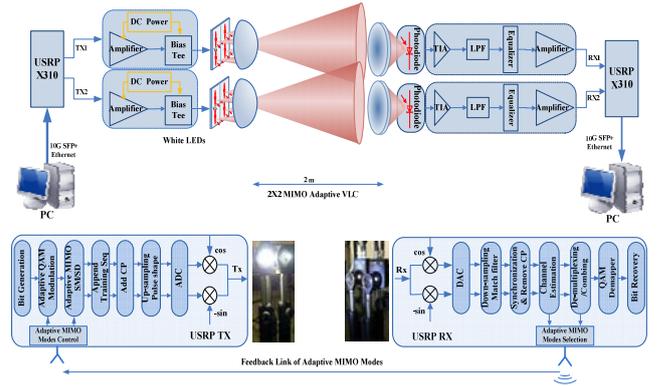

Fig. 2. Experiment setup for adaptive MIMO VLC.

IV. EXPERIMENTAL RESULTS

In order to achieve adaptive MIMO visible light communication for indoor lighting environment, we firstly evaluate efficient estimation approaches to acquire the real-time channel conditions. Then we measured and compared the error performance and spectral efficiency of $M$-QAM MIMO VLC systems using spatial diversity and spatial multiplexing, respectively. Based on the adaptive MIMO modes selection criteria in terms of channel estimation, we demonstrated the adaptive real-time software defined MIMO VLC along different distance.

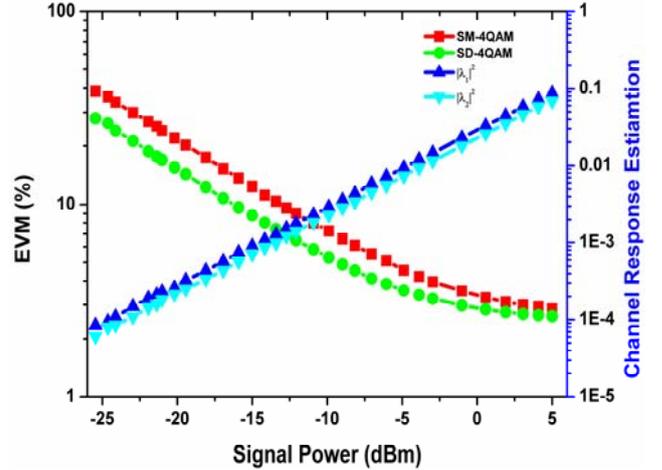

Fig. 3. EVM and Channel Estimation against signal power for 2X2 MIMO VLC using spatial multiplexing(SM) and spatial diversity(SD).

Figure 3 shows the measeured EVM and channel response estimation against signal power for 2X2 MIMO 4-QAM VLC using spatial multiplexing and spatial diversity. For large symbol streams $T$, such that $T >> N$, where $N$ is the number of unique modulation symbols, channel SNR can be obtained from EVM as $SNR \approx 1/EVM^2$. As seen in Fig. 3, EVM of SM and SD MIMO VLC decrease correspondingly with an increase in signal power. As EVM approaches to small value of 2%, both EVM become saturated due to the limited symbols streams. It's clear that EVM of spatial diversity system has a power gain of about 3dB compared to the spatial multiplexing due to the antenna array gain. Channel SNR estimation from the measured EVM is dependent on the symbols number, constellation size and MIMO processing. On the other hand, we obtain channel SNR directly from channel response estimation matrix, where $\lambda_i$ is eigenvalue of channel estimation matrix **H**. We can see in Fig.3 that each sub-channel eigenvalue of channel estimation has a linear relationship with signal power. Furthermore, original channel estimation matrix is similar for both spatial diversity and spatial multiplexing MIMO. Thus, we can acquire real-time sub-channel SNRs from eigenvalues of channel estimation matrix regardless of MIMO schemes and modulation formats.

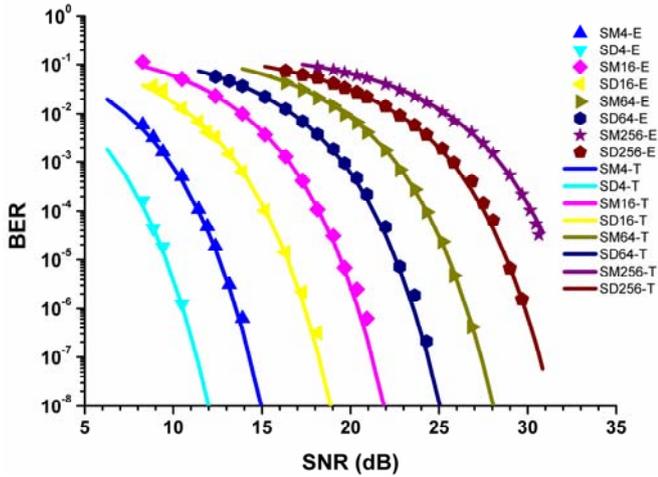

Fig. 4. BER performance of M-QAM 2X2 MIMO VLC using spatial multiplexing and spatial diversity. Dots are experimental results and lines are theoretical results.

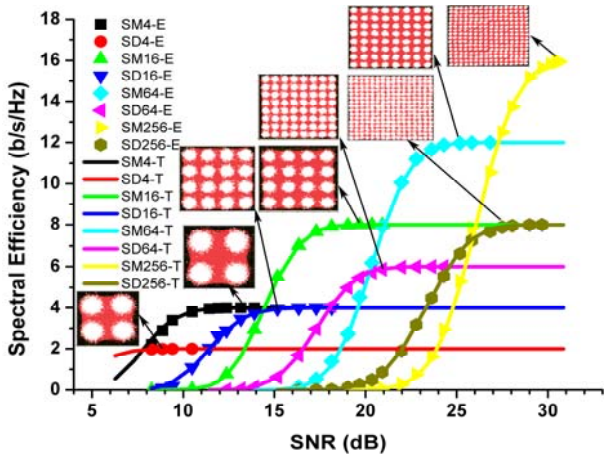

Fig. 5. Spectral efficiency performance of M-QAM 2X2 MIMO VLC using spatial multiplexing and spatial diversity. Dots are experimental results and lines are theoretical results. Insets are the corresponding constellation diagrams.

BER performance of 4-16-64-256 QAM 2X2 MIMO VLC using spatial multilpexing and spatial diversity are illustrated in Fig. 4, where dots are experimental measurements and lines are theoretical results. The results show BER decrease significantly to $10^{-7}$ as SNR grows with a linear dynamic range. Moreover, no nonlinear distortion occurs at high signal power by using single carrier $M$-QAM modulation for 2X2 MIMO VLC system. We can see that BER of spatial diversity system has about 3dB SNR gain compared to the spatial multiplexing, as the array gain in spatial diversity MIMO VLC improve the received signal strength. It is clear that the experimental measurements well match the theoretical results, which verify accurate channel SNR estimation from channel matrix in $M$-QAM MIMO VLC. Given the target BER threshold $BER_{tgt} = 10^{-3}$, we can establish the adaptive MIMO mode criteria in terms of channel SNR and singular value of channel matrix.

Figure 5 shows spectral efficiency performance of 4-16-64-256 QAM 2X2 MIMO VLC using spatial multiplexing and spatial diversity, where dots are experimental results and lines are theoretical results. Insets are the corresponding constellation diagrams. The results show spectral efficiencies of M-QAM MIMO VLC form step like shapes due to the rate discretization and BER integration. We can see that spatial multiplexing system has about 3dB gain in spectral efficiency compared with spatial diversity due to multiplexing gain of SM. These bandwidth efficiency enhancements are at the sacrifice of the degraded error performance. The experimental measurements well match the theoretical results, which verify channel SNR estimation and of spectral efficiency of $M$-QAM MIMO VLC. Thus, we can establish the adaptive MIMO mode criteria in terms of singular value of channel matrix to maximize the spectral efficiency.

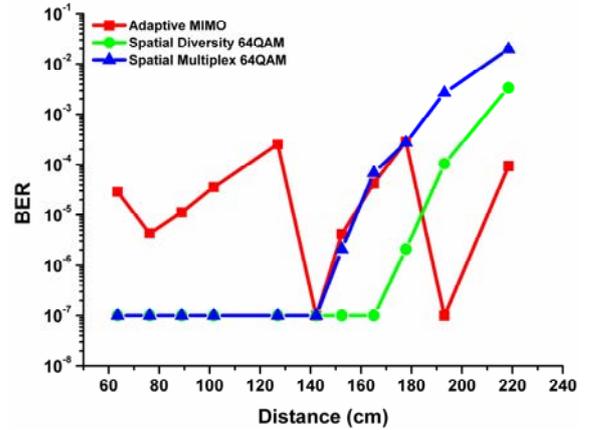

Fig. 6. BER against link distance for adaptive M-QAM MIMO, spatial multiplexing 64-QAM and spatial diversity 64-QAM VLC.

Based on the established adaptive MIMO mode criteria in terms of channel estimation matrix, we demonstrate the real-time adaptive MIMO VLC system to maximize spectral efficiency under target BER threshold. BER performance and spectral efficiency against link distance for adaptive $M$-QAM MIMO VLC are depicted in Fig. 6 and Fig. 7, respectively. The fixed 64-QAM MIMO VLC using spatial diversity and spatial multiplexing are inserted for comparison. Insets are the corresponding constellations of adaptive MIMO modes.

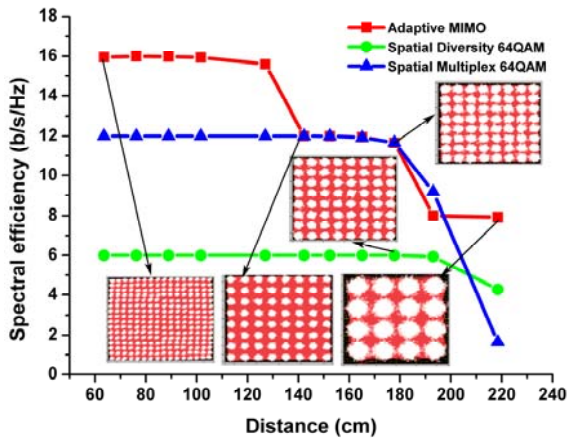

Fig. 7. Spectral efficiency against link distance for adaptive M-QAM MIMO, spatial multiplexing 64-QAM and spatial diversity 64-QAM VLC.

The initial mode is SM-64(spatial multiplex 64-QAM). As we can see in the figures, as link distance increases from 0.6 m to 1.4 m and 1.9 m, the adaptive modes change to SM-256, SM-64 and SM-16, receptively. Adaptive MIMO VLC increases the error free transmission distance to 2.2 m compared with 1.7m for SM-64 and 1.9 m for SD-64. Furthermore, adaptive MIMO VLC enhances the average spectral efficiency to 12 b/s/Hz over 2.2 m distance. Thus, it is necessary to explore the optimal combination of spatial multiplexing and spatial diversity for adaptive MIMO VLC to improve spectral efficiency and BER performance for indoor lighting environment.

## V. Conclusion

In this paper, we experimentally demonstrate a real-time software defined multiple input multiple output (MIMO) visible light communication (VLC) system employing link adaptation of spatial multiplexing and spatial diversity. Real-time MIMO signal processing is implemented by using the Field Programmable Gate Array (FPGA) based Universal Software Radio Peripheral (USRP) devices. Software defined implantation of MIMO VLC can assist in enabling an adaptive and reconfigurable communication system without hardware changes. We propose the adaptive MIMO solution that both modulation schemas and MIMO schemas are dynamically adapted to the changing channel conditions for enhancing the error performance and spectral efficiency. The average error-free spectral efficiency of adaptive 2x2 MIMO VLC achieved 12 b/s/Hz over 2 meters indoor dynamic transmission. The adaptive MIMO VLC will enhance performance in a real-time combined lighting and communication environment such as transmission power distribution, beam divergence/focusing, communication range, environmental blockage and shadowing.

## Acknowledgment

The authors would like to thank the National Science Foundation (NSF) ECCS directorate for their support of this work under Award #1201636, as well as Award #1160924, on the NSF "Center on Optical Wireless Applications".